# The infrared band strengths of $H_2O$, CO and $CO_2$ in laboratory simulations of astrophysical ice mixtures

P. A. Gerakines[1,2], W. A. Schutte[1], J. M. Greenberg[1], and E. F. van Dishoeck[1]

[1] *Leiden Observatory, P.O. Box 9513, 2300 RA Leiden, The Netherlands*
[2] *Department of Physics, Rensselaer Polytechnic Institute, Troy, NY 12180 USA*

**Abstract.** Infrared spectroscopic observations toward objects obscured by dense cloud material show that $H_2O$, CO and, likely, $CO_2$ are important constituents of interstellar ice mantles. In order to accurately calculate the column densities of these molecules, it is important to have good measurements of their infrared band strengths in astrophysical ice analogs. We present the results of laboratory experiments to determine these band strengths. Improved experimental methods, relying on simultaneous independent depositions of the molecule to be studied and of the dominating ice component, have led to accuracies better than a few percent. Furthermore, the temperature behavior of the infrared band strengths of CO and $H_2O$ are studied. In contrast with previous work, the strengths of the CO, $CO_2$, and $H_2O$ infrared features are found to depend only weakly on the composition of the ice matrix, and the reversible temperature dependence of the CO band is found to be weaker than previously measured for a mixture of CO in $H_2O$.

# 1. Introduction

Infrared astronomy has led to the identification of various species existing within icy grain mantles in dense clouds. The two most abundant molecules in grain mantles identified to date are water ($H_2O$) and carbon monoxide (CO) (e.g., Willner et al. 1982; Smith et al. 1989; Whittet et al. 1983; Whittet et al. 1985; Tielens et al. 1991; Chiar et al. 1994). Limited observational evidence as well as theoretical and laboratory modeling indicate that carbon dioxide ($CO_2$) should also be an important component of interstellar ices (d'Hendecourt & Jourdain de Muizon 1989; Whittet et al. 1989; Breukers 1991). The abundances of these molecules in different phases of ice mantles provide important clues to the chemical processes in dense interstellar clouds, and therefore it is of importance to accurately measure the band strengths of the infrared features of these molecules. The goal of this paper is to provide new, as well as more accurate measurements of the band strengths of $H_2O$, CO and $CO_2$ contained within astrophysical ice analogs, so that their abundances in the ices in dense interstellar clouds may be determined with more confidence. Such measurements are particularly important for $CO_2$, a molecule whose principle absorption band is totally obscured by Earth's atmosphere and which will be widely searched for with the Infrared Space Observatory (ISO).

Astrophysical ices consist of complex mixtures of molecules (e.g., Whittet 1992 and references therein). At least two distinct phases appear to exist: one which is dominated by polar molecules (of which the most abundant is $H_2O$) and one dominated by apolar molecules (such as CO) (e.g., Sandford et al. 1988; Tielens et al. 1991). It is not yet clear which species dominate the apolar phase, although CO and possibly $O_2$ and $CO_2$ seem to be abundant. Furthermore, observations towards embedded sources indicate a wide range of ice temperatures, from less than 20 K up to $\sim$100 K (Smith et al. 1989).

Observations of the $4.67\mu m$ ($2140\,cm^{-1}$) C$\equiv$O stretching band indicate that CO resides in both the polar and apolar phases of interstellar ice (Eiroa & Hodapp 1989; Kerr et al. 1993; Chiar et al. 1994). $CO_2$ could be present in both phases as well. Although $H_2O$ appears to be the most abundant molecule in the polar phase, the presence of apolar species like CO and $CO_2$ could influence the strength of its infrared bands. Specifically, the resulting break-up of the hydrogen bonding network would weaken the intensity of the O–H stretching feature of fully H-bonded $H_2O$ at $3\mu m$ ($3280\,cm^{-1}$) (Hagen & Tielens 1981). As of yet, $H_2O$ has not been observed in the apolar ice.

Previous methods of measuring band strengths of molecules in a mixed ice involved the preparation of several gases within one gas container by adding the components one after another and using the ideal gas law to convert pressures to abundances. This mixture is then deposited onto a cold substrate and it is assumed that the molecular abundances in the ice sample equal the gas abundances in the container. The ice abundances are then used to convert the measured integrated optical depths to infrared band strengths, either by measuring the thickness of the ice from the interference fringes produced by a laser directed at the sample and assuming a value for the ice density, thus obtaining molecular column densities (d'Hendecourt & Allamandola 1986; Hudgins et al. 1993), or by assuming that the band strengths of the dominant component are equal to those in a pure sample and using these to calibrate the other features (Sandford et al. 1988; Sandford & Allamandola 1990). These methods rely heavily on the assumed equality of the compositions of the ice sample and the gas mixture. However, several problems may arise in this assumption. First, if a bulb is made with $H_2O$ close to its vapor pressure, a small change in temperature could significantly influence the amount of $H_2O$ in the gas phase inside the bulb due to the strong temperature dependence of the vapor pressure. For example, between 18 and 23 C, it varies from 20.6 to 28.1 mbar. Second, mixing of the different gases which are sequentially allowed to enter the bulb may be incomplete. Finally, the deposition rate of a molecule will be proportional to its thermal velocity, which depends on the molecular mass, i.e., $v_{\text{th}} \propto m^{-\frac{1}{2}}$. This effect could give rise to a significant difference in the composition of the gas mixture and the ice sample if molecules with very different molecular masses are involved. For example, $H_2O$ and $CO_2$ have thermal velocities that are different by a factor of 1.56.

In the new procedure that we have implemented, we have avoided these pitfalls by producing ice mixtures using simultaneous depositions of pure gases through separate deposition tubes. We then have the ability to measure the ratio of the band strengths of molecules in a binary ice to those in pure ice ($A/A_{\text{pure}}$), since the column densities in each case (pure and mixed) can be kept the same. The band strengths in the mixed ice can then be calculated using the strengths of the pure ice bands, which are accurately known and present in the literature.

This paper is organized as follows. In § 2 we review our experimental methods. In § 3 we summarize our measurements of the band strengths of $H_2O$, CO, and $CO_2$ in various binary mixtures with polar and apolar molecules. Finally, in § 4, we discuss our results and their astrophysical implications.

## 2. Experimental

In this section we will describe the experimental equipment and procedures used for producing and analyzing astrophysical ice analogs.

### 2.1 Sample Chamber

The vacuum system used to prepare the ice samples and to obtain infrared transmission spectra is similar to those previously used to study ice analogs (Hagen et al. 1979; Allamandola et al. 1988; Hudgins et al. 1993), with some significant modifications. The chamber is situated within the sample compartment of an infrared spectrometer (Bio-Rad FTS 40A). An infrared transmitting substrate (CsI) is mounted in the vacuum chamber and can be cooled by a closed-cycle helium refrigerator (expander, Air Products Displex DE-202; compressor module, Air Products 1R04W-SL) to a temperature of ∼14 K. The temperature of the substrate is continuously adjustable by a resistive type heater element up to room temperature. The temperature is monitored by a chromel-Au thermocouple with an accuracy of 2 K. The chamber has four ports. Two of these consist of KBr, allowing transmission of the infrared beam of the spectrometer. One of the ports consists of $MgF_2$ to enable ultraviolet (UV) irradiation of the ice samples (this option was not used for the experiments described in this paper), while the fourth port consists of glass and is used for visual monitoring. Additionally, the chamber is equipped with two deposition

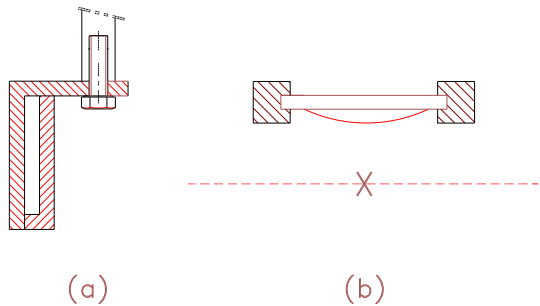

**Figure 1.** Diagram of the substrate holder, showing the 0.9 cm off-axis mounting of the substrate (a), so that it may be turned completely out of the infrared beam (b; the beam is represented by the dotted line, and the cross denotes the rotation axis)

tubes which are directed at the center of the substrate from a distance of 2.1 cm. For each deposition system, the gas flow from a storage bulb (see § 2.2) to the vacuum system is regulated by a variable leak valve (Leybold-Heraeus 28341). The leak valve is equipped with a shut-off valve and a regulation valve that function independently. The stainless-steel chamber is sealed with Viton O-rings and is evacuated by a turbo-molecular pump (Pfeiffer/Balzers TPH170), backed by a rotary pump (Pfeiffer/Balzers DUO016B). A liquid nitrogen trap between the two pumps prevents any rotary pump oil from backstreaming to the system. Ion and thermocouple pressure gauges placed between the sample chamber and the turbo pump monitor the internal pressure. The vacuum system can be externally heated using heating tape. After about two days of pumping and externally heating to $\sim$80 C, a vacuum of $1 \times 10^{-7}$ mbar was obtained. The residual gases at this pressure were analyzed by collecting them on the substrate after cooling to 14 K and subsequent infrared spectroscopy. It was found to consist mainly of $H_2O$, accreting at a rate of $\sim 2 \times 10^{12}$ molec cm$^{-2}$ s$^{-1}$ (0.002$\mu$m hr$^{-1}$), while organic contaminants were found to accrete at about half this rate.

The substrate holder can be rotated without breaking the vacuum. During an experiment, the substrate is rotated between two positions which are 90 degrees apart. In pos. 1, an infrared transmission spectrum of the substrate and the ice sample can be obtained. The spot size of the infrared beam on the CsI substrate is 0.6 cm in diameter. Position 2 enables deposition of an ice sample through one or both deposition tubes onto the cold substrate. Also, the ice sample can be UV-irradiated in this position. Moreover, the substrate holder is constructed such that the substrate is mounted 0.9 cm away from the rotation axis. Therefore, while the substrate is in pos. 2, the infrared beam may pass unimpeded through the sample chamber (see Fig. 1). This allows the collection of a reference infrared spectrum at any moment during the experiment. Use of such reference spectra greatly improves the baseline stability of the infrared spectra (see § 2.4).

The homogeneity of the ice samples produced with the two deposition tubes was checked by varying the spot size of the infrared beam on the sample using the variable aperture in front of the infrared source of the spectrometer. Samples were found to be homogeneous within 10% inside a diameter of 1.2 cm.

## 2.2 Gas Bulb Preparation

Bulbs for individual gases were prepared in a glass vacuum manifold, evacuated through an oil diffusion pump (Edwards Diffstack Series Model 63) backed by a rotary mechanical pump (Edwards BS2212). Back streaming of pump oil into the system is prevented by a liquid nitrogen cold trap located between the line and the pumps. The pressure in the line can be monitored in many ranges with an ion gauge, a thermocouple gauge, and a diaphragm manometer. The system is heated with heater tape (under vacuum) to clean it between preparation of different samples. The compounds used and their purities are as follows: $H_2O$ (liquid), triply distilled; CO (gas), Messer Griesheim, 99.997% purity; $CO_2$ (gas), medical grade, 99.5% purity; $O_2$ (gas), Messer Griesheim, 99.998% purity.

Pressures used in bulbs of $H_2O$ were limited to a maximum of 10 mbar, well below its room temperature vapor pressure of $\sim$ 20 mbar. We have thus avoided the effects of substantial sticking of $H_2O$ on the walls of the bulb and the temperature dependence of the $H_2O$ vapor pressure, as described in § 1.

## 2.3 Experimental Procedures

The experimental procedures applied for measuring the infrared band strengths for a molecule in a binary ice are described here. Two gas bulbs, one containing the gas for which the infrared band strengths are to be measured (henceforth the "subject" gas) and one containing the gas with which this species is to be diluted (henceforth the "dilutant") are connected to the entries of the two deposition tubes. Before cooling the substrate, flow rates are set with the regulation valve while monitoring the pressure increase within the system, using the following relation:

$$F_i \propto \frac{\Delta P}{\sqrt{m_i}} , \qquad (1)$$

where $m_i$ is the molecular mass of component $i$, $\Delta P$ is the measured pressure increase within the vacuum system, and $F_i$ is the desired flow rate of component $i$ (Schutte et al. 1993). The flow rates through each deposition tube may be set independently in this way based on the desired ratio of the component gases. After setting the flow rates with the regulation valve, flows are controlled exclusively with the shut-off valve. From later spectroscopic analysis (as described in § 2.4) of ices produced by this method of precalibration, it is found that the flow rates reproduce within 3%.

After calibration, the substrate is cooled down to 14 K, and a deposition is made of the subject gas. After obtaining the infrared spectrum, the substrate is heated until the sample sublimes and then recooled. Next, the subject gas and the dilutant are deposited simultaneously for the same length of time as the first deposition, resulting in a binary ice sample containing the same number of subject molecules. The column density $N$ of an ice component and the integrated optical depth of its absorption band $i$ relate as:

$$N = \frac{\int \tau_i(\nu)d\nu}{A_i} , \qquad (2)$$

where $A_i$ is the band strength and the integration is taken over the feature in question using an appropriate baseline. Thus, the ratio of the strengths in the binary and in the pure ice can be found simply from the ratio of the integrated optical depths in the two samples. If the ratio of subject:dilutant is large ($<$ about 10%), an additional sample of pure dilutant is made as a final step. Then the exact composition of the binary ice can be determined from the integrated optical depths measured in the two pure samples using band strengths from the literature. If the subject material makes up less than 10% of the binary ice, the composition of the binary ice is simply determined from the integrated optical depths of the dilutant in the binary ice and of the subject material in the pure ice, and no additional depositions are performed.

Ice samples used in our experiments had the approximate composition dilutant:subject = 20:1 or 1:1. Due to uncertainties involved in pre-calibrating the flows using the induced pressure increase, the actual composition of the ice sample as it was measured from the infrared spectra could differ by up to a factor of two from what was intended, although the deviation was typically no more than $\sim$30%. The total deposition rate when producing the binary ice was typically $\sim 4 \times 10^{15}$ molec cm$^{-2}$ s$^{-1}$ for the 20:1 ices and $\sim 1 \times 10^{15}$ molec cm$^{-2}$ s$^{-1}$ for the 1:1 ices (corresponding to thickness growth rates of $\sim$4 and 1$\mu$m hr$^{-1}$, respectively). Deposition rates of the subject gases were $\sim 2 \times 10^{14}$ molec cm$^{-2}$ s$^{-1}$ when producing a 20:1 ice sample and $\sim 5 \times 10^{14}$ molec cm$^{-2}$ s$^{-1}$ when producing a 1:1 ice sample. With these deposition rates, the contamination due to mixing of the residual gases into the ice samples ranged from $\sim$0.1 to 2%. Deposition times were equal to 6 min for the 1:1 ices and 3 min for the 20:1 ices, and hence the resulting thicknesses of the ice samples were $\sim$0.2 and 0.1$\mu$m respectively. Depositions at 20 K instead of 14 K were made in order to check the sticking of the molecules on the substrate, i.e., if no decrease in sticking at the higher substrate temperature is found, then sticking can be assumed to equal 100% (Sandford & Allamandola 1990). In all cases, sticking was found to equal unity.

## 2.4 Measurement of Spectra

The sample chamber is situated within the sample compartment of the infrared spectrometer such that the infrared beam axis is aligned with the center of the cold substrate in pos. 1. Single-beam spectra were taken from 4400 to 400 cm$^{-1}$ (2.3 to 25$\mu$m) at a resolution of 1.0 cm$^{-1}$ (the width of an unresolved line). Producing a sample spectrum involved obtaining a single-beam spectrum of 100 scans before and after deposition, with subsequent ratioing (see e.g., Hudgins et al. 1993).

In some cases, especially when measuring spectra after gradual warm-up of the ice sample, the time between obtaining the background and the sample may become large–perhaps on the order of hours. Due to inevitable spectrometer instabilities, such as in the source temperature and in the alignment of the interferometer, this can lead to the appearance of some spurious spectral structure and deviation of the spectral baseline from zero absorbance. To prevent this, additional single-beam spectra can be obtained just after the background spectrum (at time $t_1$) and just after the sample spectrum (at time $t_2$) with the substrate in pos. 2, i.e., with the infrared beam passing unimpeded through the vacuum chamber (an "empty" spectrum). The sample absorbance spectrum is then obtained from:

$$\text{Abs} = -\log_{10}\left[\frac{\text{Sample}(t_2)}{\text{Empty}(t_2)}\right] + \log_{10}\left[\frac{\text{Background}(t_1)}{\text{Empty}(t_1)}\right] . \qquad (3)$$

## 3. Results

### 3.1 Pure ices

Integrated optical depths (in cm$^{-1}$) were measured from the absorbance spectra by choosing an appropriate baseline for the band in question. For bands of pure ice and for most ices without H$_2$O, a linear baseline was used. The bands of H$_2$O are strong and broad, and they dominate in most of the mixed ice spectra. Most CO$_2$ and CO bands lie on top of these broad structures. In order to define a good baseline in these cases, we have made polynomial fits to the underlying structure and used those which closely approximate its curvature. In general, good fits were obtained for polynomials of order 2 or 3. In some cases, an average of measurements was taken of a band for which more than one polynomial fit well. An example of a band for which the underlying H$_2$O feature created some difficulty in the determination of the baseline is the CO$_2$ 660 cm$^{-1}$ (15$\mu$m) band, which lies on top of the H$_2$O 13$\mu$m (760 cm$^{-1}$) feature. This band and two polynomial fits of the underlying H$_2$O feature are shown in Figure 2 for an H$_2$O:CO$_2$ = 24:1 mixture. The difference in measured integrated optical depth between the two cases is 14%.

Table 1 lists the infrared band strengths for the pure ices used in these experiments. For each molecule, values for the strongest features have been taken from the literature (Yamada & Person 1964; Jiang et al. 1975; Hagen et al. 1981) and used to calculate the strengths for the other bands by scaling the relative integrated optical depths. For the bands of $^{13}$CO and $^{13}$CO$_2$, a terrestrial isotopic ratio of $^{12}$C/$^{13}$C = 89 has been used to further scale the band strengths. Measurements of pure CO$_2$ at higher temperatures were obtained by depositing the ice at 14 K, followed by stepwise heating of the ice sample at a rate of about 2 K min$^{-1}$ and taking an infrared spectrum at each step. This procedure reveals only a weak temperature dependence of the CO$_2$ infrared bands. Band positions in both $\mu$m and cm$^{-1}$ are listed for the pure ices (cf. Sandford et al. 1988; Sandford & Allamandola 1990). For purposes of identification, we will denote each band by its position in the pure ice. Peak band positions in mixed ices were found to agree with those of Sandford et al. (1988) and Sandford & Allamandola (1990), and we refer the reader to these works for a complete listing of band positions in binary mixtures.

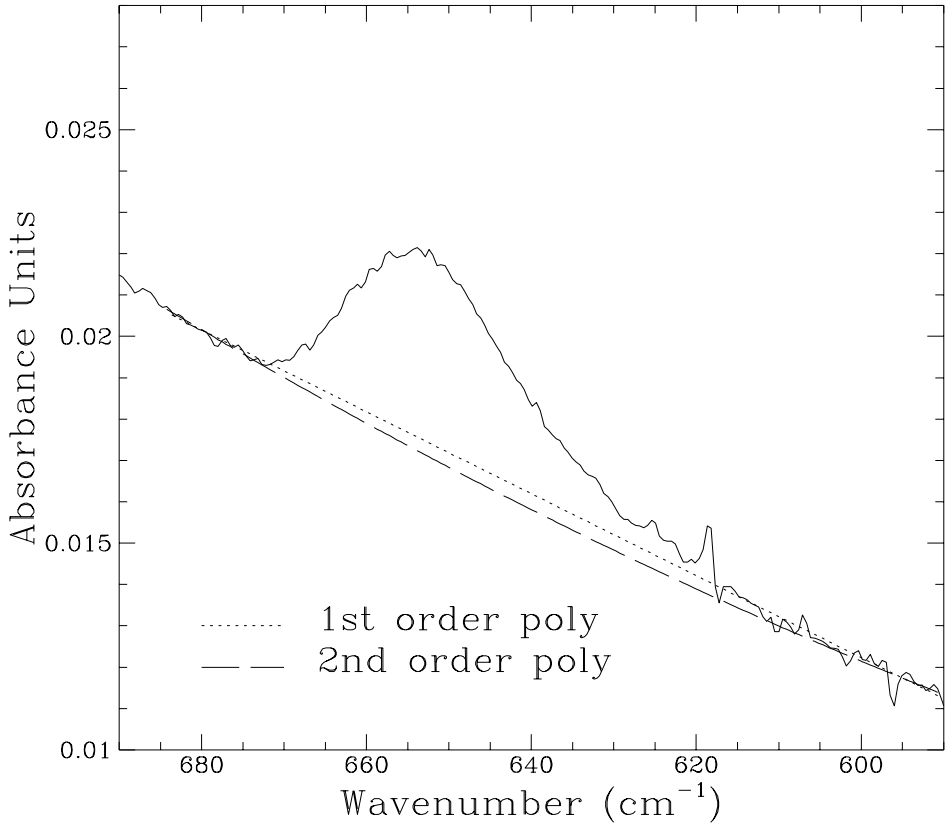

**Figure 2.** The $660\,\text{cm}^{-1}$ ($15\mu\text{m}$) band of $CO_2$ in an $H_2O:CO_2 = 24:1$ mixture demonstrates the uncertainty involved in producing a baseline fit due to the underlying feature of $H_2O$. Dotted line– 1st order polynomial fit to the $H_2O$ band; dashed line– 2nd order fit in the same region

## 3.2 CO mixtures

The band strengths for the $^{13}C\equiv O$ and $^{12}C\equiv O$ fundamental stretching modes are shown in Table 2, for CO mixed with $H_2O$, $O_2$, and $CO_2$. Errorbars were estimated from the results obtained with various polynomials used as a baseline fit. Only errors larger than 5% are listed. Errors due to spectral noise are negligible (i.e., less than 1%), even for the weakest features. In each case, the value of the $^{12}CO$ band intensity ratioed by that of pure CO does not deviate from unity by more than 13%. The values for the $^{13}CO$ band in $H_2O$ contain a substantial baseline error due to its small size and its position on a broad $H_2O$ feature. Within the errorbars, these values remain close to the pure ice value as well. This also holds for non-$H_2O$ ices, since the value of the $^{13}CO$ band strength appears to stay close to the value for the pure ice within the given uncertainty. It must be noted that since $O_2$ is infrared-inactive, the composition of the $O_2$ containing binary ices could only be assessed from the increase in pressure at room temperature by the $O_2$ flow (§ 2.3). Thus, for these ices there can be an error in the listed composition of about a factor of 2

**Table 1.** Infrared band intensities of pure $H_2O$, CO, and $CO_2$ ices after deposition at 14K, and $CO_2$ after warm-up to 60 and 100 K

| Ice | Mode | Band Position cm$^{-1}$ ($\mu$m) | $A_{14K}$ cm molec$^{-1}$ | $A_{60K}/A_{14K}$ | $A_{60K}$ cm molec$^{-1}$ | $A_{100K}/A_{14K}$ | $A_{100K}$ cm molec$^{-1}$ |
|---|---|---|---|---|---|---|---|
| $H_2O$ | O–H stretch | 3280 (3.045) | 2.0(-16)[a] | | | | |
| | O–H bend | 1660 (6.024) | 1.2(-17) | | | | |
| | libration | 760 (13.16) | 3.1(-17) | | | | |
| CO | $^{12}C\equiv O$ stretch | 2139 (4.675) | 1.1(-17)[b] | | | | |
| | $^{13}C\equiv O$ stretch | 2092 (4.780) | 1.3(-17) | | | | |
| $CO_2$ | ($\nu_3$) $^{12}C=O$ stretch | 2343 (4.268) | 7.6(-17)[c] | 0.98 | 7.4(-17) | 0.97 | 7.4(-17) |
| | ($\nu_3$) $^{13}C=O$ stretch | 2283 (4.380) | 7.8(-17) | 0.94 | 7.3(-17) | 0.92 | 7.2(-17) |
| | ($\nu_2$) O=C=O bend | 660,665 (15.15,15.27) | 1.1(-17) | 1.03 | 1.1(-17) | 1.04 | 1.1(-17) |
| | ($\nu_1 + \nu_3$) combination | 3708 (2.697) | 1.4(-18) | 1.05 | 1.5(-18) | 1.08 | 1.5(-18) |
| | ($2\nu_2 + \nu_3$) combination | 3600 (2.778) | 4.5(-19) | 1.22 | 5.5(-19) | 1.20 | 5.4(-19) |

a(-b)=a$\times 10^{-b}$; [a] Hagen et al. 1981; [b] Jiang et al. 1975; [c] Yamada & Person 1964

(of course, this does not introduce any uncertainty in the measurement of $A/A_{\mathrm{pure}}$, see § 2).

We have studied the temperature dependence of the $^{12}C\equiv O$ band strength in the $H_2O$:CO = 26:1 ice sample. Measuring the irreversible component of the temperature dependence is impossible for CO, since the weakly-bound fraction of the CO molecules already starts to sublime when the sample temperature is raised to 25 K (Schmitt et al. 1989; Sandford et al. 1988). For this reason, we have only measured the reversible component of the temperature dependence, following the method of Schmitt et al. (1989).

The sample was initially deposited at 14 K and then heated to 115 K and allowed to anneal for 20 minutes to sublime any weakly-bound CO from the sample. The sample was then held at a temperature of 100 K for 5 minutes, in order to stop sublimation and to pump away any gas-phase CO left in the system. The sample is then recooled and measured at 14 K. After annealing, the ice composition was found to equal $H_2O$:CO = 30:1, when assuming that A(CO,14 K) = $1.1\times10^{-17}$ cm molec$^{-1}$, as for the unannealed ice. As a check of CO re-condensation during this cooldown, the sample was kept at 14 K for 10 min and then re-measured. No new condensation was found in our ice sample. The sample is then warmed up in steps, and a spectrum taken at each step. In order to check whether any further sublimation of CO occurred during this second warm-up sequence, the sample was once again cooled to 14 K, and the CO band measured. The intensity of the CO band was found to remain constant relative to the initial annealed ice. Figure 3 shows the obtained reversible temperature dependence of the $^{12}C\equiv O$ fundamental stretching mode. As in Schmitt Et al. (1989), a clear temperature dependence is found. Our results show that the band strength drops by 17% when the temperature is raised from 14 to 90 K.

**Table 2.** Infrared band intensities of CO in binary ices with $H_2O$, $O_2$ and $CO_2$ after deposition at 14K. Uncertainties greater than 5% are listed

| Ice | Mode | $A/A_\text{pure}$ | $A$ cm molec$^{-1}$ |
|---|---|---|---|
| $H_2O$:CO=26:1 | $^{12}C\equiv O$ stretch | 0.97 | 1.1(-17) |
| | $^{13}C\equiv O$ stretch | 0.63±0.3 | (8.2±4)(-18)[a] |
| $H_2O$:CO=2.1:1 | $^{12}C\equiv O$ stretch | 0.99 | 1.1(-17) |
| | $^{13}C\equiv O$ stretch | 0.78±0.3 | (1.0±0.4)(-17)[a] |
| $O_2$:CO=20:1[b] | $^{12}C\equiv O$ stretch | 1.13 | 1.2(-17) |
| | $^{13}C\equiv O$ stretch | 1.18±0.1 | (1.5±0.1)(-17) |
| $O_2$:CO=1:1[b] | $^{12}C\equiv O$ stretch | 1.04 | 1.1(-17) |
| | $^{13}C\equiv O$ stretch | 1.00 | 1.3(-17) |
| $CO_2$:CO=17:1 | $^{12}C\equiv O$ stretch | 1.04 | 1.1(-17) |
| | $^{13}C\equiv O$ stretch | 0.87±0.1 | (1.1±0.1)(-17) |
| $CO_2$:CO=0.8:1 | $^{12}C\equiv O$ stretch | 1.10 | 1.2(-17) |
| | $^{13}C\equiv O$ stretch | 1.04 | 1.4(-17) |

a(-b) = a×10$^{-b}$; [a] Measurement of integrated optical depth contains a large uncertainty due to underlying features of $H_2O$; [b] Listed composition is uncertain by a factor of 2 (see text)

### 3.3 $CO_2$ mixtures

Table 3 presents the band strengths for the infrared bands of $^{12}CO_2$ and the strongest band of $^{13}CO_2$ in $H_2O$, $O_2$, and CO mixtures. Errors due to baseline uncertainties are listed whenever they exceed 5%, as in Table 2. We have investigated the temperature dependence of the $CO_2$ bands in an $H_2O$ matrix by warm-up of the sample following deposition, as discussed above for pure $CO_2$ (see §3.1). No significant temperature dependence was found (at 100 K, the strengths of all $CO_2$ bands measured deviated by less than 3 % from their values at 14 K). As with CO, the $CO_2$ band strengths depend only slightly on the ice composition. The strength of the main stretching feature of $CO_2$ varies by less than 10% for the ices used here. The strengths of the $^{13}CO_2$ band and the bending mode of $CO_2$ seem to deviate by a somewhat larger amount (up to 21 and 37% respectively), but these measurements can contain considerable error (up to 20%). Only the strengths of the two weak $CO_2$ combination bands definitely show considerable dependence on the ice composition.

In order to estimate the influence of residual gases depositing within the ice samples, an $H_2O$:$CO_2$ = 20:1 experiment was performed with a 4 times higher deposition velocity.

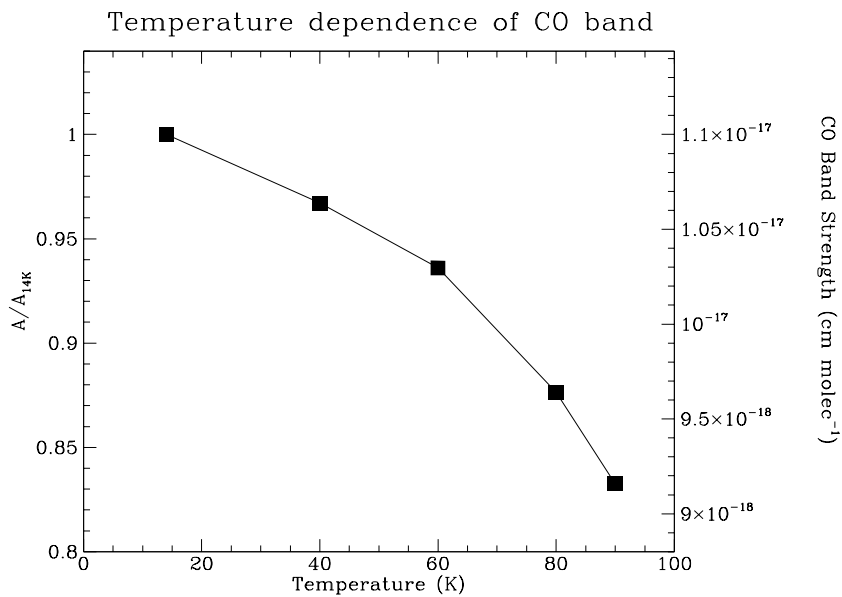

**Figure 3.** The reversible temperature dependence of the CO infrared band strength in an $H_2O$ dominated matrix, $H_2O:CO=30:1$. The absolute scale shown on the right y-axis is obtained by assuming $A(CO,14K) = 1.1\times10^{-17}$ cm molec$^{-1}$, as in the ice sample before annealing

The obtained pure to binary ratios deviated by less than 1% from those obtained at the lower deposition rate, indicating that the residual gas condensation does not affect the results.

### 3.4 $H_2O$ mixtures

Figures 4, 5, and 6 show the effects of diluting $H_2O$ with CO and $CO_2$ on the 3, 6, and 13$\mu$m (3280, 1660, and 760 cm$^{-1}$) $H_2O$ bands, respectively. The 3$\mu$m band could be measured with a simple straight baseline for all cases, but for the binary ices, the side bands produced were subsequently subtracted using a polynomial fit (of order 2 or 3) to the wing of the 3$\mu$m band as a baseline. The resultant correction was less than 2.5% of the total band intensity in all cases. For the 6$\mu$m band, we used a straight baseline through the regions 1050 - 1080 cm$^{-1}$ and 1930 - 1980 cm$^{-1}$, and for the 13$\mu$m band a straight baseline from 1050 to 500 cm$^{-1}$ was used. For the $H_2O:CO_2 = 1.6:1$ mixture, the $CO_2$ O=C=O bending mode at 15$\mu$m (660 cm$^{-1}$) totally obscures the shape of the long-wavelength side of the $H_2O$ 13$\mu$m band. In this case, the $H_2O$ feature was assumed to be symmetrical, and twice the measured integrated optical depth from its peak frequency ($\sim 770$ cm$^{-1}$, depending on temperature) to 1050 cm$^{-1}$ was taken as an estimate of its full integrated optical depth.

For both CO and $CO_2$, the 3$\mu$m band is slightly reduced in strength after initial deposition at 14 K. As the ice is heated, however, these molecules begin to diffuse through

**Table 3.** Infrared band intensities of $CO_2$ in binary ices with $H_2O$, $O_2$, and CO after deposition at 14 K. Uncertainties greater than 5% are listed

| Ice | Mode | $A/A_{\mathrm{pure}}$ | $A$ cm molec$^{-1}$ |
|---|---|---|---|
| $H_2O$:$CO_2$=24:1 | ($\nu_3$) $^{12}$C=O stretch | 0.94 | 7.1(-17) |
|  | ($\nu_3$) $^{13}$C=O stretch | 0.80±0.1 | (6.2±0.8)(-17) |
|  | ($\nu_2$) O=C=O bend | 1.38±0.2 | (1.5±0.2)(-17) |
|  | ($\nu_1 + \nu_3$) combination | ... | ...[a] |
|  | ($2\nu_2 + \nu_3$) combination | ... | ...[a] |
| $H_2O$:$CO_2$=1.6:1 | ($\nu_3$) $^{12}$C=O stretch | 1.00 | 7.6(-17) |
|  | ($\nu_3$) $^{13}$C=O stretch | 0.98 | 7.6(-17) |
|  | ($\nu_2$) O=C=O bend | 1.37±0.2 | (1.5±0.2)(-17) |
|  | ($\nu_1 + \nu_3$) combination | 0.83±0.2 | (1.2±0.3)(-18) |
|  | ($2\nu_2 + \nu_3$) combination | 0.72±0.3 | (3.2±1)(-19) |
| $O_2$:$CO_2$=20:1[b] | ($\nu_3$) $^{12}$C=O stretch | 0.95 | 7.2(-17) |
|  | ($\nu_3$) $^{13}$C=O stretch | 0.79±0.1 | (6.2±0.8)(-17) |
|  | ($\nu_2$) O=C=O bend | 0.85 | 9.4(-18) |
|  | ($\nu_1 + \nu_3$) combination | 0.85 | 1.2(-18) |
|  | ($2\nu_2 + \nu_3$) combination | 1.02 | 4.6(-19) |
| $O_2$:$CO_2$=1:1[b] | ($\nu_3$) $^{12}$C=O stretch | 1.05 | 8.0(-17) |
|  | ($\nu_3$) $^{13}$C=O stretch | 1.10 | 8.6(-17) |
|  | ($\nu_2$) O=C=O bend | 0.91 | 1.0(-17) |
|  | ($\nu_1 + \nu_3$) combination | 1.00±0.1 | (1.4±0.1)(-18) |
|  | ($2\nu_2 + \nu_3$) combination | 1.36 | 6.1(-19) |
| CO:$CO_2$=29:1 | ($\nu_3$) $^{12}$C=O stretch | 1.09 | 8.3(-17) |
|  | ($\nu_3$) $^{13}$C=O stretch | 1.00 | 7.8(-17) |
|  | ($\nu_2$) O=C=O bend | 0.96 | 1.1(-17) |
|  | ($\nu_1 + \nu_3$) combination | 2.15 | 3.0(-18) |
|  | ($2\nu_2 + \nu_3$) combination | 1.58 | 7.1(-19) |
| CO:$CO_2$=1.3:1 | ($\nu_3$) $^{12}$C=O stretch | 1.01 | 7.7(-17) |
|  | ($\nu_3$) $^{13}$C=O stretch | 0.99 | 7.7(-19) |
|  | ($\nu_2$) O=C=O bend | 0.88 | 9.7(-18) |
|  | ($\nu_1 + \nu_3$) combination | 1.04±0.1 | (1.5±0.1)(-18) |
|  | ($2\nu_2 + \nu_3$) combination | 1.33±0.2 | (6.0±1)(-19) |

a(-b) = a×10$^{-b}$; [a] Band is unobservable due to overlap with strong $H_2O$ band; [b] Listed composition is uncertain by a factor of two (see text)

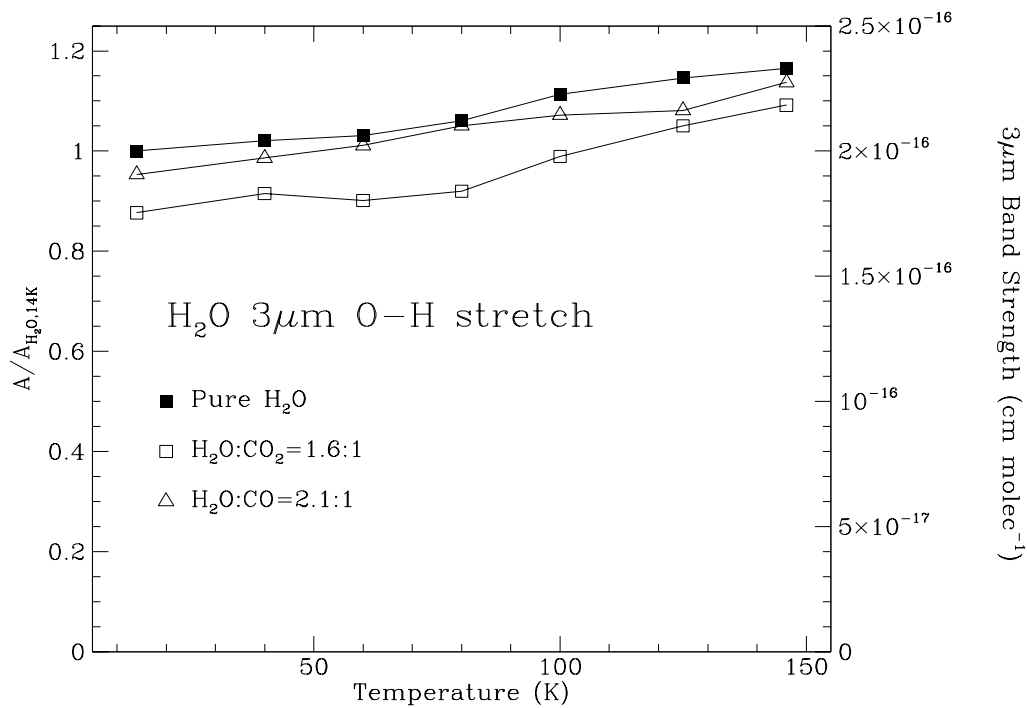

**Figure 4.** Measured values of the integrated optical depth of the $H_2O$ $3\mu m$ O–H stretching feature as a function of temperature in different mixtures, ratioed by the integrated optical depth of the $3\mu m$ band in pure $H_2O$ directly after deposition at 14 K. Solid squares– pure $H_2O$; empty squares– $H_2O:CO_2$=1.6:1; empty triangles– $H_2O:CO$=2.1:1

and escape from the $H_2O$ matrix (Schmitt et al. 1989; Sandford et al. 1988), and the $3\mu m$ band grows as more $H_2O$ molecules form hydrogen bonds. Just after deposition at 14 K, the $6\mu m$ band is slightly strengthened, and the $13\mu m$ band is slightly weakened by the new molecule. Both bands approach the pure $H_2O$ value as they are heated.

## 4. Discussion

### 4.1 Comparison with previous studies

The results of our experiments show that the band strengths of the infrared absorption features of CO and $CO_2$ depend only weakly on the composition of the ice mixture in which they are diluted. Overall, while the infrared absorption features of molecules may change considerably in width depending on the dilutant molecule (Sandford et al. 1988; Sandford & Allamandola 1990), our results show that their band strengths remain quite constant, i.e., an increase in width is compensated by a decrease in depth. For example, although the CO band near $2140\,\mathrm{cm}^{-1}$ is more than three times wider in an $H_2O$ dominated ice than in pure CO (Sandford et al. 1988), the band strength only varies by a few percent. It has also been shown that the strengths of the 3, 6, and 13$\mu$m (3280, 1660, and 760 $\mathrm{cm}^{-1}$) bands of $H_2O$ are only weakly affected by the presence of CO and $CO_2$, i.e., the change is < about 20% for all three features for ices where the amount of apolar species is comparable to the amount of $H_2O$.

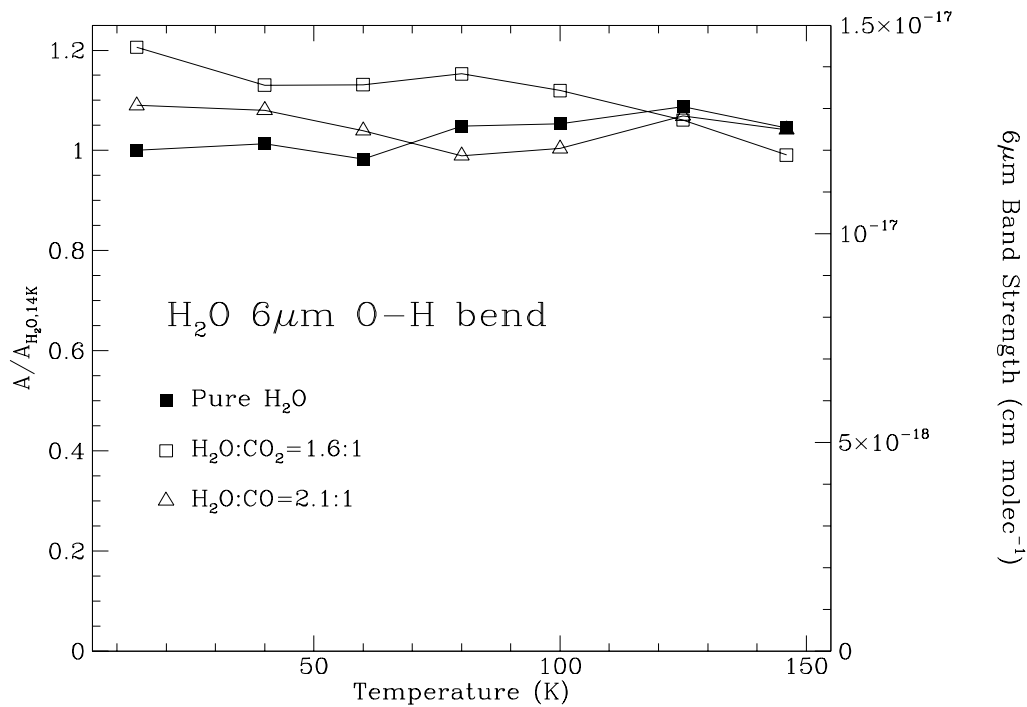

**Figure 5.** As Fig. 4, but for the $H_2O$ 6$\mu$m O–H bending mode

The band strengths determined for CO and $CO_2$ diluted in $H_2O$, $O_2$, CO and $CO_2$ stay very close to the values for the pure ices. This is in contrast with earlier studies which have shown large increases in the band strengths of these molecules when diluted in $H_2O$, i.e., by 70% for CO and by factors of 2 to 4 for the features of $CO_2$ (Sandford et al. 1988; Palumbo & Strazulla 1993; Sandford & Allamandola 1990). Since the errors in our results are in general less than 10%, it appears that these earlier measurements could be significantly influenced by the problems involved in the deposition method as discussed in § 1. It must be noted that band strength measurements made with these earlier methods for other molecules also indicate strong variations in different matrices, e.g, in the case of $CH_4$ (Hudgins et al. 1993). On the other hand, the band strengths of $CO_2$ in a CO matrix do agree well with those measured previously (Sandford & Allamandola 1990). Additional careful studies should be performed to study band strengths for Other astrophysically interesting molecules.

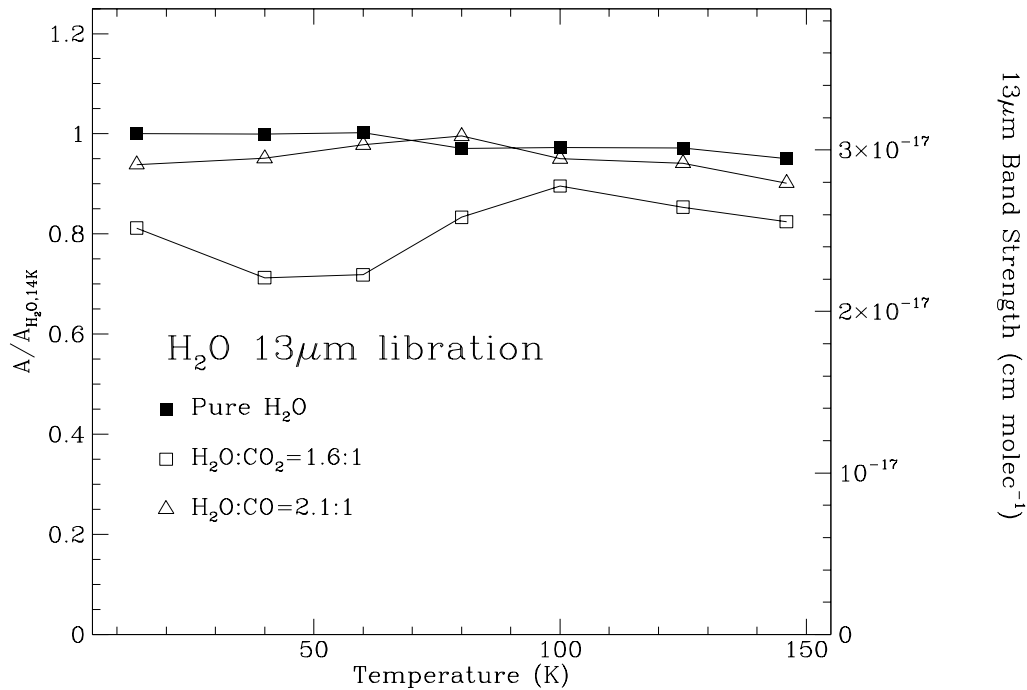

**Figure 6.** As Fig. 4, but for the $H_2O$ 13$\mu$m libration mode

It is found that the 2140 cm$^{-1}$ band of CO in an $H_2O$:CO ice shows a reversible dependence on temperature, decreasing by 17% if the temperature is raised from 14 K to 90 K. Earlier measurements showed a considerably larger decrease of 33% over this temperature range (Schmitt et al. 1989). The difference between these results may be attributed to the use of specular reflectance for obtaining infrared spectra in this earlier study, since it has been shown that measurements made with this technique give different peak absorbances of infrared features of ice samples as compared to spectra measured in transmission (Kitta & Krätschmer 1983) due to interference losses at the interface between the ice sample and the reflecting block surface (Hagen et al. 1981).

When $H_2O$ is diluted with other molecules, the strengths of its absorption bands are affected, as discussed in § 3.4 and shown in Figs 4, 5 and 6. We find that the strength of the O–H stretching band at 3$\mu$m in a mixture of $H_2O$:$CO_2$ = 1.6:1 is reduced to 88% of its value in a pure $H_2O$ ice and to 95% in an $H_2O$:CO = 2.1:1 mixture (see Fig. 4). The effects of dilution on the 3$\mu$m $H_2O$ absorption band have been studied previously by Greenberg et al. (1983). Using the statistical concentrations of monomers, dimers, and trimers of a molecule randomly placed within a simple cubic lattice (Behringer 1958), Greenberg et al. (1983) have shown that this band will only be present for mixtures with an $H_2O$ concentration above 15% and that the band strength of the 3$\mu$m feature will approximately be related to the $H_2O$ concentration $f$ by the semi-empirical relation

$$\frac{A_{\text{mix}}}{A_{\text{pure}}} = \frac{1 - f_0/f}{1 - f_0} , \qquad (4)$$

where $f_0 = 0.15$ is the lower limit for producing the $3\mu m$ band. With $f = 0.62$ and $0.68$, for the $H_2O:CO_2$=1.6:1 and $H_2O:CO$=2.1:1 mixtures, Eq.(4) yields $A_{\text{mix}}/A_{\text{pure}} = 0.89$ and 0.92, respectively, which are close to the measured reductions of 88% and 95%. The consistency of these results may indicate that there is good mixing of the subject and dilutant molecules and that there is little diffusion taking place upon deposition before the molecules become fixed in the ice lattice.

## 4.2 Astrophysical implications

Our results have several astrophysical implications. First, the abundance of CO in the polar or $H_2O$ dominated phase of icy grain mantles obtained from the infrared spectra of ices in dense interstellar clouds becomes 1.7 times larger than previously calculated. Toward embedded sources in dense clouds, the amount of CO in polar ice is then comparable to the amount of CO in apolar ice (Tielens et al. 1991; Chiar et al. 1994). Also, the $CO_2$ column densities as calculated by d'Hendecourt & Jourdain de Muizon (1989) toward three sources become 3.7 times higher, matching or even exceeding that of CO towards the objects studied. Next, ices toward embedded sources have been observed with line-of-sight averaged temperatures up to $\sim 80$ K (Smith et al. 1989). In this case, the band strength applied to determine the column density of CO in polar ice should be taken slightly lower than the value of $1.1 \times 10^{-17}$ cm molec$^{-1}$ measured at 14 K. For example, for an average ice temperature of 70 K, a CO band strength of $1.0 \times 10^{-17}$ cm molec$^{-1}$ is more suitable (Fig. 3). Finally, in determining $H_2O$ column densities from the 3 and $6\mu m$ interstellar absorption bands, the band strengths for pure $H_2O$ ice may be used, since the presence of small abundances of apolar species ($<$ about 30%) in the ice has little influence on its band strengths. However, the effects of dilution should become important if the concentration of apolar molecules would be considerably larger, i.e., $\geq 60\%$ Eq.(4), as noted by Greenberg et al. (1983).

**Ackowledgements.** We would like to acknowledge fruitful discussions with Peter Jenniskens. This work was partially funded by NASA grant NGR 33-018-148.